\documentclass[12pt]{article}

\usepackage{amsmath,amssymb,graphicx,mathrsfs,slashed,setspace}

\usepackage[title]{appendix}
\usepackage{float}
\usepackage{latexsym}
\usepackage{bm}
\usepackage{blindtext}
\usepackage{hyperref}
\hypersetup{
	colorlinks=true,
	linkcolor=blue,
	filecolor=black,
	urlcolor=black,
	citecolor=blue,
}

\newcommand{\hide}[1]{}

\newcommand{\be}{\begin{equation}}
	\newcommand{\ee}{\end{equation}}
\newcommand{\bea}{\begin{eqnarray}}
	\newcommand{\eea}{\end{eqnarray}}

\def\({\left(}
\def\){\right)}

\begin{document}
	\title{\vspace{-1.8in}
		{Defrosting the Born-Infeld dyonic frozen star with tachyon matter: spectrum of oscillations}}
	\author{\large Ram Brustein${}^{(1)}$,  A.J.M. Medved${}^{(2,3)}$, Tamar Simhon${}^{(1)}$
		\\
		\vspace{-.5in} \hspace{-1.5in} \vbox{
			\begin{flushleft}
				$^{\textrm{\normalsize
						(1)\ Department of Physics, Ben-Gurion University,
						Beer-Sheva 84105, Israel}}$
				$^{\textrm{\normalsize (2)\ Department of Physics \& Electronics, Rhodes University,
						Grahamstown 6140, South Africa}}$
				$^{\textrm{\normalsize (3)\ National Institute for Theoretical Physics (NITheP), Western Cape 7602,
						South Africa}}$
				\\ \small \hspace{0.57in}
				ramyb@bgu.ac.il,\ j.medved@ru.ac.za,\ simhot@post.bgu.ac.il
			\end{flushleft}
	}}
	\date{}
	\maketitle

	\begin{abstract}

The frozen star is a model for the interior of an astrophysical black hole that is externally indistinguishable from the singular geometry of a Schwarzschild black hole, even though it contains no singularities nor trapped surfaces. The frozen star corresponds to a macroscopic ``BIon'', a solution of a certain type of Born-Infeld Lagrangian coupled to Einstein gravity,
which consists of perfectly rigid flux tubes sourced by an electric charge localized at its center and a uniform distribution of
opposite charge along its outer surface. This configuration is a solution of the effective action that describes the end point of tachyon condensation in string theory when the tachyon is at the minimum of its potential. The rigidity of the flux tubes implies that the star is ultrastable under linear perturbations of the geometry and  matter. The star can be effectively  deformed, or ``defrosted'', to allow for some non-trivial internal pulsations, as expected from regular and horizonless astrophysical black holes. To describe the  defrosted star, we find that it is necessary to include a perturbative amount of magnetic-monopole charge and tachyon kinetic energy. This recasts the star as a  macroscopically large ``DIon'', whose flux tubes can stretch and contract, having  both electric and magnetic charges localized at its center and distributed uniformly along its outer surface. Previously, the non-radial oscillations of the defrosted star were analyzed, to leading order in a perturbative ``defrosting'' parameter $\gamma$, by describing the energy-momentum tensor as that of fluid modes. Here, we derive the same spectrum of oscillating modes in a much simpler way, using the Born-Infeld Lagrangian. Thus, we verify that they  exhibit non-relativistic sound velocities scaling as $\gamma$ and possess parametrically long lifetimes scaling as $1/\gamma^2$. Because the perturbation equations are derived  from a Lagrangian,  the consistency of their dynamics is guaranteed for arbitrary deviations away from spherical symmetry, facilitating numerical studies on frozen stars away from their  equilibrium configuration.

		\end{abstract}
		\newpage
		\setstretch{1.5}

		\renewcommand{\baselinestretch}{1.5}
		\section{Introduction}

		While the astrophysical evidence for black holes (BHs) is robust, their standard description is marred by the paradoxes
		that are inherent to any spacetime exhibiting  geodesic incompleteness and signature-flipping horizons \cite{FW}. As such, many have turned to alternative models of ultracompact but regular objects as a means of dispelling with these paradoxical inconsistencies \cite{carded,BHmimickers,Carballo-Rubio:2025fnc}. These so-called BH mimickers are designed to
		eliminate singularities and trapped surfaces, while maintaining observational indistinguishability from
		general-relativistic BHs to external observers, at least at the level of equilibrium physics.

A realization of just such a BH mimicker  that has so far passed every test is the frozen star model \cite{bookdill,BHfollies}. Subsequent elaborations   include \cite{popstar,trajectory,fluctuations,U4Euclidean,Tom1} and a generalization to include rotation can  be found in \cite{notstevekerr}. The model has recently been improved to include an explicit description of its  source of matter \cite{StringFluid}, which itself has been further studied  in \cite{Tunnel} and generalized to include rotation in \cite{StringPlusRot}.

The matter source in our improved model  is described by the Lagrangian of a string fluid  that is formally reminiscent to that of a Born--Infeld theory  \cite{GHY,Cookie,Yi}.  This Lagrangian describes the end point of the decay of an unstable $D$-brane or a brane–antibrane system via a process of  open-string tachyon condensation, as described by  Sen \cite{Sen,Sen2,Sen3,Sen4,Sen5}. Adopting  the recasted, Born--Infeld  form of Sen's Lagrangian in \cite{GHY,Cookie,Yi} and then coupling it to gravity, we obtained a  static, spherically symmetric solution with a distinct internal structure; namely, a set of perfectly rigid, radially oriented tubes of electric flux that form a connection between a point-like central source and an outer, spherical shell of equal and opposite charge. Thus, the frozen star can be viewed as a gravitationally confined BIon \cite{Gibbons:1997xz,GibbonsRev} because,  in our case,  the gravitational
          backreaction terminates the  flux lines at the boundary of
          the object.

          The frozen star is meant as an effective classical (or even semiclassical)  incarnation of the collapsed polymer BH \cite{inny,strungout,emerge}, which models  the interior of an ultracompact object as a highly excited fluid of closed strings and defies having a classical geometric description.

          An integral property of the frozen star is its ultrastability, which is a direct consequence of the equation of state
          $\;p+\rho=0\;$, where $p$ is the radial component of pressure
          and $\rho$ is the energy density. This stability insures that any perturbation of a  matter density or of an element
          of a geometric tensor
          is identically vanishing \cite{bookdill,popstar} and is analogous to that of the polymer model, which is ``hairless'' (in the BH sense of the word) in
          the limit that the string coupling $g_s$ or, equivalently, $\hbar$ goes to zero.

For finite $g_s$, the polymer BH is known to support non-radial oscillations whose spectrum is characterized by a  sound velocity that scales with $g_s$  and a lifetime that scales with $1/g_s^2$ \cite{collision}. The key point here is that the string coupling  is now small but finite, $\;0<g_s \ll 1\;$. Meaning  that, if one is to investigate the oscillatory spectrum of  a frozen star, it is necessary to ``defrost" the star by assuming small departures from its classical state of  maximally negative radial pressure. These deviations can be quantified by a positive, dimensionless parameter $\gamma$, which must be chosen to be sufficiently small to apply a perturbative treatment but should still be large enough to  dominate over an  exponentially small parameter $\varepsilon^2$, which determines how much $|g_{tt}|$ or $g^{rr}$ deviates from zero at the location of the would-be horizon,  $\;\varepsilon^2\ll \gamma \ll 1\;$.

This notion of a defrosted star was first implemented  in \cite{fluctuations}, where the Cowling approximation \cite{Cowling}, which assumes the complete decoupling of the matter and geometric fluctuations, was applied to determine the non-radial oscillation spectrum for  fluid perturbations in a defrosted  star  background. This framework was further developed in \cite{Tom1}, where  the coupling between fluid modes and even-parity metric perturbations was also incorporated. In both cases, the results indicate that these non-radial oscillations behave just like those of a  polymer BH, given that the defrosting parameter $\gamma$ has been swapped in for $g_s$. The sound velocities of the modes are non-relativistic, scaling as $\gamma$, and have parametrically long lifetimes, scaling as $1/\gamma^2$.  The latter is  a distinct observational signature that contrasts sharply with the rapidly decaying quasi-normal modes of standard BHs.

In our previous articles, we described the matter in terms of a fluid energy--momentum tensor, which could only  be defined consistently for a spherically symmetric solution.  In this paper, we describe the matter in terms of a Lagrangian, which guarantees the consistency of the perturbation equations, even  for departures away from spherically symmetric configurations. We are then able to validate our previous results by a much simpler analysis within the framework of the improved model. To achieve this, however,  it is necessary to include in the Lagrangian  some magnetic-monopole  charges and tachyon matter in addition to the electric charges. The resulting macroscopically large ``DIon'' \cite{Shwinger-Dyons} features both an electric point charge and a magnetic monopole at its center, along with a respective pair of equal and oppositely charged layers that are distributed uniformly over the outermost surface of the star. As a result of this setup, there is an interior Dirac string \cite{Dirac1} that runs along a localized portion of the  polar axis and  connects the central monopole to the layer of magnetic charge  at the outer surface. The tachyon matter allows the flux tubes to contract and stretch, in contrast to their perfect rigidity in its absence.

The rest of the paper is arranged as follows: The next three sections provide a brief review of the  formalism that is needed to follow the main analysis. These are, in their presented order, discussions on the frozen star geometry, the defrosted star framework and the Born--Infeld matter source.  The novel analysis begins in Section~5, where  the Born--Infeld formalism is suitably modified for the defrosted star generalization; in particular, for the explicit inclusion of  magnetic-monopole charge and tachyon matter. We discuss the resulting Dirac string in the  gravitationally confined model in Section~6.  The perturbation equations of the defrosted star geometry and Born--Infeld gauge fields are presented and then solved in Section~7.
The closing  Section~8 contains a brief overview and an appendix
justifies a condition that is imposed in Section~7.


		\section{The frozen star solution}

Along with the standard assumptions of spherical symmetry and
a static spacetime, a key feature of the frozen star geometry is that the bulk~\footnote{The precise meaning of the ``bulk'' in this context will be made clear below.}  of the star's interior is
sourced by matter with a maximally negative radial pressure, $\;p=-\rho\;$.
The transverse pressure, $q$, vanishes throughout the bulk of the star as a consequence of the condition $\;p+\rho=0\;$ being imposed on  the Einstein equations and  Einstein-tensor conservation equation. Importantly, it is this same  condition
that is responsible for the frozen star's complete rigidity.

Another key feature is that every spherical slice of the bulk interior
is exponentially close to being a null surface. With that in mind,
the interior metric  for the frozen star then takes the form
		\be
		ds^2\;=\; -\varepsilon^2 dt^2+ \frac{1}{\varepsilon^2} dr^2+ r^2d\Omega^2\;,
		\label{FSmetric}
		\ee
where the dimensionless parameter $\varepsilon^2$ should be regarded
as exponentially small.

The fluid densities describing this geometry go as
\bea
\label{polyrho}
8\pi G  \rho &=& \frac{1-(rf)'}{r^2} \;=\;  \frac{1-\varepsilon}{r^2}\;, \\
\label{polypr}
8\pi G  p &=& -\frac{1-(rf)'}{r^2} \;=\; -\frac{1-\varepsilon}{r^2}\;, \\
\label{polypt}
8\pi G  q  &=& \frac{(rf)''}{2r}\;=\; 0\;,
\eea
where $\;f=|g_{tt}|\;$, $\;g^{rr}=f\;$~\footnote{The second equality is a consequence of $\;p=-\rho\;$ and not assumed.} and a prime denotes a radial derivative, here and throughout the paper.

Meanwhile,
the conservation equation reduces from its general static, spherically symmetric  form of
		$\; p'+ \frac{1}{2}(\ln{f})'\left(\rho+p\right)+\frac{2}{r}\left(p-q\right)=0\;$ into
		\be
		q\; = \;\frac{1}{2 r} \partial_r(r^2 p)\;.
		\label{conserv2}
		\ee

It is necessary to modify the metric in Eq.~(\ref{FSmetric})
 over a thin transitional
layer close to (and at) the outermost surface  so that the star's metric and its first two derivatives
can be matched in a  smooth manner to the exterior Schwarzschild metric \cite{popstar}. It is also necessary to regularize
the geometry close to (and at) the center of the star  so
that the energy density remains finite as the solution
approaches $\;r=0\;$ \cite{trajectory}.

The width of the transitional surface layer $\lambda$  should be regarded as small, on the order of the string or Planck length scale. Hence,  its inclusion in calculations would lead to highly suppressed corrections.
The transverse pressure, though, grows very
large in the layer, $\;q\sim \frac{\rho}{\lambda}\;$, as can be understood by an inspection of Eq.~(\ref{conserv2}).
Similarly, the regularized core has a just-as-small width and
can similarly be ignored in our analysis.

			\section{The defrosted star solution}

		Because of the ultrastability of the frozen star solution, modifications are required to mimic the quantum-induced oscillatory modes of the polymer BH. In particular, there must  be a deviation away from the equation of state $\;p=-\rho\;$ --- which necessitates $\;|g_{tt}| \neq g^{rr}\;$ --- and at least one of $g_{tt}$ and $g^{rr}$ has to become radially dependent.

		To implement these changes formally, we introduce a positive, dimensionless parameter $\gamma$, such that
		$\;\varepsilon^2 \ll \gamma \ll 1\;$.
		To leading order, the relevant metric components are then expressed as
		\begin{align}
			-g_{tt} &= \gamma\left(\frac{r}{R}\right)^a\;, \\
			g^{rr} &= \gamma\left(\frac{r}{R}\right)^b\;,
		\end{align}
		whereby the null-energy condition stipulates that $\;a\geq b\;$.
		Moreover, the  self-consistency of
		the resulting set of linearized equations imposed the specific
		choices of
		$\;a=2\;$ and $\;b=0\;$ \cite{fluctuations}.

		Using Einstein's equations, we have for  the density  and radial pressure $p$,
		\begin{align}
			8\pi G \;r^2\; \rho &\;=\; 1-(1+b)\gamma\left(\frac{r}{R}\right)^b\;=\; 1-\gamma \;,
\label{TmunurDef}
			\\
		8\pi G \;	r^2 \; p &\;=\; -1+(1+a)\gamma\left(\frac{r}{R}\right)^b\;=\;-1+3\gamma\;,
\label{TmunupDef}
\end{align}
		which can be combined to give
		\begin{equation}
			8\pi Gr^2 (\rho+p) \;=\; (a-b)\gamma\left(\frac{r}{R}\right)^b\;=\; 2\gamma\;,
\label{TmunuqDef}
\end{equation}
		for which the deviation away from the ultrastable equation of state $p=-\rho$, is  manifestly clear.

		Substituting the above densities into the conservation equation, we find for the transverse pressure that
		\begin{equation}
			q \;=\; \frac{1}{4}\left(\frac{a^2+ab+2b}{a-b}\right)(\rho+p)\;=\; \frac{1}{2}(\rho+p)\;.
		\end{equation}

		Matching $g^{rr}$ to the exterior Schwarzschild metric,~\footnote{For the purposes of this estimate, we are disregarding the transitional layer.}
		 one observes that  the defrosted star's radius $R$ is slightly larger than its frozen star counterpart, which is exponentially close to the star's Schwarzschild radius,
		\begin{equation}
			R \;\simeq\; 2GM(1+\gamma)\;>\; 2GM(1+\varepsilon^2) \;\gtrsim\; 2GM\;,
		\end{equation}
		where $M$ is the star's mass.

		In spite of  this increased radius, one finds that integrating the energy density over the volume leads (at leading perturbative order)
		to the same mass $M$. This implies that that a frozen star
		of mass $M$ and its associated  set of  defrosted stars, as labeled by the perturbative parameter $\gamma$,  form  a family of (nearly) degenerate states.


		\section{The Born-Infeld matter source}

		In this section, we review
		the pertinent aspects of including the Born--Infeld
		Lagrangian of \cite{StringFluid} in the frozen star framework.
		This matter Lagrangian was first presented and discussed
		in \cite{GHY} (also see \cite{Cookie,Yi}).
		Let us take note of some conventions:
		An index of 0 denotes time, indices $a,b,\dots$ denote spacetime dimensions  and $i,j,\dots$ denote spatial dimensions.  Four-dimensional spherical coordinates
		will be employed, such that  $\;(0,1,2,3)=(t,r,\theta,\phi)\;$.

		 The starting  point of this analysis is
		 Sen’s effective action for the decay of unstable $D$-branes at the conclusion  of tachyon condensation \cite{Sen,Sen2,Sen3,Sen4,Sen5}. Near the vanishing minimum of the tachyon potential (or $D$-brane tension), the  effective Lagrangian (density)
		 can be expressed as~\footnote{We are neglecting the kinetic term for each of  the scalar fields representing
		 a transverse dimension and, for the moment, that of  the tachyon field; see \cite{Yi} for further details.}
		\begin{equation}
			{\cal L} \;= \;- V(T)\sqrt{-\text{Det}\left(\eta+ 2\pi \alpha'{\cal F}\right)} + \sqrt{-\eta} A^a J_a \;.
\label{TachLag1}
		\end{equation}
		 Here,  $V(T)$ represents the  potential for the tachyon $T$, $2\pi \alpha'$ is the inverse of the fundamental string tension, $\eta^{a}_{\;\;b}=\delta^{a}_{\;\;b}$ is the Minkowski background metric and ${\cal F}_{ab}=\partial_{a}A_{b}-\partial_{b}A_{a}$ is the field-strength tensor for the gauge field $A_a$ with source $J_a$.

Although this Lagrangian vanishes when $V(T)$ vanishes, the Hamiltonian density  ${\cal H}$, remains well-defined. In the absence of magnetic sources, this is
		\begin{equation}
			{\cal H} \;=\; D^i{\cal F}_{0i}-{\cal L} \;=\; E_i D^i \;,
		\end{equation}
		where the final equality follows from  $\;{\cal L}=0\;$ at the minimum of the potential. Here, $\;E_i={\cal F}_{0i}\;$ is the electric field and $D^i$ is the electric displacement which  is the canonical conjugate of the gauge field, $\;D^i=\delta {\cal L}/\delta (\partial_0 A_i)\;$, and is the fundamental electric field in a Born–Infeld theory because it satisfies the Gauss’-law constraint $\;\partial_i D^i = \sqrt{-\eta} J_0 = \sqrt{-\eta} \rho_e\;$.

		More generally, the Hamiltonian takes the form $\;{\cal H} = \frac{1}{2\pi\alpha'}\sqrt{D^i D_i + {\cal P}^i{\cal P}_{i}}\;$, where $\;{\cal P}_i = ({\vec{D}\times\vec{B}})_i\;$ is the conserved momentum that is associated  with spatial translations and $\vec{B}$ is the magnetic induction.   As always, one includes charges by  imposing the Gauss’-law constraint, Ampere’s law and the Bianchi identities (including $\;\partial_i B^i = 0$).

Following \cite{GHY}, which adopts ideas from \cite{Tsey}, we  define a  nonvanishing dual Lagrangian by regarding the field-strength elements ${\cal F}_{ij}$ as conjugates with respect to a new set of dual variables which are  defined as  $\;K^{ij}=2\delta {\cal H} / \delta {\cal F}_{ij}\;$. The dual  Lagrangian is then  defined by a suitable Legendre transformation,
		\begin{equation}
			{\cal L}' \;=\; {\cal H}-\frac{1}{2}{\cal F}_{ij}K^{ij} \;=\; \frac{1}{2\pi\alpha'}\sqrt{D^iD_i-\frac{1}{2}K^{ij}K_{ij}}\;.
		\end{equation}

		Again guided by \cite{GHY}, we introduce a new gauge field $\widetilde{A}_a$ and use it to  define an effective field strength  $\;{\cal K}_{ab}=\partial_a \widetilde{A}_b- \partial_b \widetilde{A}_a\;$, using the relations $\;{\cal K}_{0i}= D_i\;$ and $\;{\cal K}_{ij}= K_{ij}\;$ or,   equivalently,
		$\;{\cal K} = D_1 dt\wedge dx^1  + \frac{1}{2} K_{ij} dx^i\wedge dx^j\;$.
		Then, with the help of  $\;\frac{1}{2}K^{ij}K_{ij} = \frac{1}{{\cal H}^2}{\cal P}^{i}{\cal P}_{i}D^j D_j\;$, the new Lagrangian
		can be recast into an explicitly Born--Infeld form,
		\begin{equation}
			{\cal L}' \;=\; \frac{1}{2\pi \alpha'}\sqrt{-\frac{1}{2}{\cal K}^{ab}{\cal K}_{ab}}\;.
		\end{equation}
		In the case of the absence of magnetic sources, this new Lagrangian reduces to the original Hamiltonian, $\;{\cal L}' = E_iD^i\;$.

		The field strengths  ${\cal F}_{ab}$ and ${\cal K}_{ab}$ are to  be regarded as independent variables, a distinction that is relevant for deriving the equations of motion. Equivalently, the two gauge fields, the original one $A_a$ and the new one $\widetilde{A}_a$,  should be  viewed as independent.

		A subtle point about this procedure is that the definition of ${\cal K}_{ab}$ implies that $\;{\cal K}\wedge {\cal K}=0\;$,
		the imposition of which  requires  a Lagrange-multiplier term to be added to the Lagrangian; otherwise, the    canonical analysis of ${\cal L}'$ will not lead back  to the correct form of ${\cal H}$ \cite{GHY}. With this addition, the complete Born–-Infeld Lagrangian
		is
		\begin{equation}
			{\cal L}' \;=\; \frac{1}{2\pi \alpha'}\sqrt{-\frac{1}{2}{\cal K}^{ab}{\cal K}_{ab}} + \lambda_1 \epsilon^{abcd}{\cal K}_{ab}{\cal K}_{cd} + \sqrt{-g} A^aJ_a\label{lag1}\;,
		\end{equation}
		where $\lambda_1$ is the Lagrange multiplier and we have allowed for the  subsequent generalization to curved space by taking $\;\sqrt{-\eta}\to \sqrt{-g}\;$.

When this Lagrangian is coupled to gravity, another Lagrange-multiplier term is required to ensure that the total mass is a conserved quantity, so that the star's radius is fixed.  This second Lagrange-multiplier term provides an effective repulsive force that exactly  cancels the attractive electrostatic force between the  oppositely charged distributions on the gravitationally  confined BIon, thus stabilizing	the overall configuration \cite{StringFluid}. In any event, both types of Lagrange-multiplier terms vanish once their respective constraints are imposed, thus neither contributes to the equations of motion.

		The two-form ${\cal K}^{ab}$ can also be  identified as a surface-forming bivector, leading to  its  interpretation as a cross-sectional slice of the world sheet of an open string or, pictorially, as a tube  of electric flux \cite{GHY}. The equations of motion for this Lagrangian, $\;d{\cal L}'=0\;$,  are equal to the original Bianchi identities $\;d{\cal F}=0\;$. Conversely,  the Bianchi identities for the new field strength $\;d{\cal K}=0\;$ are the same as  the equations of motion  of the original Lagrangian ${\cal L}$. Moreover, although $A_a$ is not the gauge field for ${\cal K}_{ab}$, the variation of ${\cal L}'$ by $A^a$ does lead to  the correct Gauss’-law constraint for the displacement field $D^i$.

		As justified in \cite{Letel}, the energy--momentum tensor is given, when away from any source,  by $\;T_{ab} = 2\delta {\cal L}'/\delta g^{ab}\;$, which yields
		\begin{equation}
			T_{ab} \;=\; \frac{1}{2\pi\alpha'}\frac{{\cal K}_{a}^{\;\;c}{\cal K}_{bc}}{\sqrt{-\frac{1}{2}{\cal K}^{de}{\cal K}_{de}}}\;,
			\label{stressed}
		\end{equation}
		translating into  $\;T_{00} = {\cal H}\;$, $\;T_{0i} = -{\cal P}_i\;$, and $\;T_{ij} = \frac{1}{2\pi\alpha'}(-D_iD_j+{\cal P}_i{\cal P}_j)/{\cal H}\;$.

		The simplest solution to the equations of motion is a static, spherically symmetric configuration where $\;{\cal K}_{01}=-{\cal K}_{10}\;$ and all other elements vanish. In which case, $\;{\cal K}^{ab}{\cal K}_{ab}=-2 D_1 D^1\;$ and the only nonvanishing elements of the energy--momentum tensor are
	\begin{equation}
			T_{00} \;=\; -T_{11} = \frac{1}{2\pi\alpha'}\sqrt{D^1D_1}\;,
		\end{equation}
		which  implies an equation of state of  $\;p=-\rho\;$ and $\;q=0\;$. This configuration is known as the ``string fluid'' in \cite{GHY} and   is intimately related to the BIon solution in \cite{Gibbons:1997xz} given that   $\;D_1 \propto 1/r^2\;$, corresponding to a point-like charge at the origin. Therefore, when decoupled from gravity, BIons can be viewed as being sourced by a fluid of electric flux lines emanating radially from a central point source and extending to infinity.

		Importantly for us, this same static, spherically symmetric configuration  with   $\;D_1 \propto 1/r^2\;$ also describes
		the bulk interior of a frozen star. The difference between our
		case and the standard BIon is that, for ours,  there is  an equal and opposite distribution of charge on the outer surface of the compactified BIon
		where the electric flux lines must then terminate.
		As discussed above, a Lagrange-multiplier constraint for conserved mass ensures the stability of this net-neutral  configuration of separated charges.

		\section{The Born-Infeld source for a defrosted star}

To connect the energy--momentum tensor of Eq.~(\ref{stressed})  to that of  a defrosted star, for which  $\;q\propto (\rho+p)\propto \gamma\;$,  will require some additional work.   To obtain a transverse pressure, the new ingredient has to be a magnetic field, but one that  does {\em not}  break spherical symmetry; namely,  one that is sourced by  a magnetic monopole.

To better understand this,  first consider that any mixing of a 0 index or a 1 index  with a 2 index and/or a 3 index in the contraction  ${\cal K}^{ab}{\cal K}_{ab}$ would lead to an off-diagonal term ({\em i.e.}, a spherical-symmetry breaking term)  in the energy--momentum tensor. Hence, to preserve spherical symmetry, the only other non-vanishing elements besides
$\;{\cal K}_{01}=-{\cal K}_{10}\;$ could be $\;{\cal K}_{23}=-{\cal K}_{32\;}$, which would necessarily describe a magnetic field in the radial direction. For this inclusion to be realized, we simply extend  the previous definition of ${\cal K}_{ab}$  into
\be
{\cal K}\;=\; D_1 dt\wedge dx^1 \;+\;  B_1 dx^2\wedge dx^3 \;+\;
\frac{1}{2} K_{ij} dx^i\wedge dx^j\;,
\label{calK1}
		\ee
where $B_1$ is meant to be the magnetic analogue of $D_1$.

The effective field-strength tensor for this more general but still spherically symmetric configuration can then be expressed as
\be
{\cal K}^{ab}{\cal K}_{ab}\;=\; -2\left[D^iD_i - B^iB_i\right]_{\delta_{i=1}}\;.
\ee
The relative sign difference is due to the first term being contracted with the time--time element of the metric, whereas the second term  only involves the angular elements of the metric.

The nonzero (diagonal) energy--momentum tensor elements  now take  the form,
		\bea
		T^{0}_{\;\;0}\;= \;T^{1}_{\;\;1} &=& -\frac{1}{2\pi\alpha'}\frac{D^i D_i}{\sqrt{D^jD_j -B^jB_j}}\;, \label{stressedasef}\\
		T^{2}_{\;\;2} \;=\; T^{3}_{\;\;3} &=& \frac{1}{2\pi\alpha'}\frac{B^iB_i}{\sqrt{D^jD_j -B^jB_j}}\;.
		\label{stressedasf}
		\eea
It can be  observed that  $\;\rho+p =0\;$, which needs to be modified for any solution of the defrosted star.

To obtain a solution with  $\;\rho+p > 0\;$, an additional source is needed,  tachyon matter, which is indeed included in the framework of $D$-brane decay \cite{SenReview}. In this case,  the Lagrangian in Eq.~(\ref{TachLag1}) is modified by adding a  kinetic  term for the scalar tachyon field $T$ to the Lagrangian \cite{Yi,GHY},
\begin{equation}
			{\cal L} \;= \;- V(T)\sqrt{-\text{Det}\left(\eta+2\pi \alpha'{\cal F}
+\partial_\mu T \partial_\nu T  \right)} + \sqrt{-\eta} A^a J_a \;.
\label{TachLag2}
		\end{equation}

We next  consider a homogeneous tachyon background, same  as in \cite{GHY,Cookie,Yi}, such that $\;\partial_i T=0\;$ but $\;\partial_0 T \ne 0\;$.
This amounts to adding an energy density for the tachyon matter along the  direction of the  flux.

It is  possible to define a modified antisymmetric tensor by adding on to Eq.~(\ref{calK1})  a  fifth (dummy) coordinate that represents the tachyonic degree of freedom
\cite{Cookie},
\be
		{\cal K}\;=\; D_1 dt\wedge dx^1 \;+\;  B_1 dx^2\wedge dx^3 \;+\; \frac{1}{2} K_{ij} dx^i\wedge dx^j+\pi_T dt\wedge dT\;,
\label{calK2}
		\ee
such that $\;D_T=0\;$ and $\pi_T$, the tachyon conjugate momentum, is the temporal component of a 4-vector.
In the case of the homogeneous tachyon matter,  the Lagrangian ${\cal L}'$ in Eq.~(\ref{lag1}) is still
\begin{equation}
			{\cal L}' \;=\; \frac{1}{2\pi \alpha'}\sqrt{-\frac{1}{2}{\cal K}^{ab} {\cal K}_{ab}}\;,
		\end{equation}
but now  with $\;a,~b = 0,1,2,3,T\;$.

The modified energy--momentum tensor elements in terms of contractions of  $\mathcal{K}_{ab}$ differ, formally,   from those in terms of
$D_i$ and $B_i$ fields in Eqs.~(\ref{stressedasef},\ref{stressedasf}) only by the presence of a
$\mathcal{K}^{T0}\mathcal{K}_{0T}$ term in the numerator
of $T^0_{\;\;0}$,
		\begin{eqnarray}
				T^0_{\;\;0} &=& \frac{1}{2\pi\alpha'}\frac{\mathcal{K}^{10}\mathcal{K}_{10}+\mathcal{K}^{T0}\mathcal{K}_{T0}} {\sqrt{-\frac{1}{2}{\cal K}^{ab} {\cal K}_{ab}}}\;, \nonumber
				\\
				T^1_{\;\; 1}&=&\frac{1}{2\pi\alpha'}\frac{\mathcal{K}^{01}\mathcal{K}_{01}} {\sqrt{-\frac{1}{2}{\cal K}^{ab} {\cal K}_{ab}}}\;,
				\nonumber
\\				T^2_{\;\;2}&=&T^3_{\;\;3}\;=\;\frac{1}{2\pi\alpha'}\frac{\mathcal{K}^{23}\mathcal{K}_{23}} {\sqrt{-\frac{1}{2}{\cal K}^{ab} {\cal K}_{ab}}}\;.
			\label{Tmunu-BI}
		\end{eqnarray}
Note that $\;-\rho=T^0_{\;\;0}<0\;$, $\;p=T^1_{\;\;1}<0\;$ and $\;q=T^2_{\;\;2}>0=T^3_{\;\;3}>0\;$  and  that the denominators have also been implicitly modified,
\be
\sqrt{-\frac{1}{2}{\cal K}^{ab} {\cal K}_{ab}}\; =\; \sqrt{|\mathcal{K}_{01}|^2+|\mathcal{K}_{0T}|^2-|\mathcal{K}_{23}|^2}\;,
\label{denom}
\ee
where the notation $|\mathcal{K}_{AB}|^2$ means $\frac{1}{2}|\mathcal{K}^{ab}\mathcal{K}_{ab}|_{a,b=\{A,B\}}\;$.   Also, $\;D^iD_i=|\mathcal{K}_{01}|^2\;$ and $\;B^iB_i=|\mathcal{K}_{23}|^2\;$.

\subsection{Calculating $\mathcal{K}_{ab}$ for the defrosted star}

Comparing Eqs.~(\ref{Tmunu-BI}) to Eqs.~(\ref{TmunurDef}-\ref{TmunuqDef}), we obtain

\begin{eqnarray}
\frac{1-\gamma}{8\pi G r^2} &=&\frac{1}{2\pi\alpha'}\frac{|\mathcal{K}_{01}|^2+|\mathcal{K}_{0T}|^2} {\sqrt{-\frac{1}{2}{\cal K}^{ab} {\cal K}_{ab}}}\;, \nonumber
				\\
\frac{1-3 \gamma}{8\pi G r^2}
&=&\frac{1}{2\pi\alpha'}\frac{|\mathcal{K}_{01}|^2} {\sqrt{-\frac{1}{2}{\cal K}^{ab} {\cal K}_{ab}}}\;,
				\nonumber
\\
\frac{\gamma}{8\pi G r^2}&=&\frac{1}{2\pi\alpha'}\frac{|\mathcal{K}_{23}|^2} {\sqrt{-\frac{1}{2}{\cal K}^{ab} {\cal K}_{ab}}}\;,
			\label{Tmunu-BI-Def}
		\end{eqnarray}
with $\sqrt{-\frac{1}{2}{\cal K}^{ab} {\cal K}_{ab}}$ given by Eq.~(\ref{denom}).
Solving the three equations, we find that, to leading order in $\gamma$,
\bea
|\mathcal{K}_{01}|^2 &=& (1- 5\gamma) \left(\frac{2\pi \alpha'}{8\pi G r^2}\right)^2\;,
\cr
|\mathcal{K}_{0T}|^2&=& 2\gamma \left(\frac{2\pi \alpha'}{8\pi G r^2}\right)^2\;,
\cr
|\mathcal{K}_{23}|^2&=&\gamma \left(\frac{2\pi \alpha'}{8\pi G r^2}\right)^2\;,
\label{soldef}
\eea
as well as
\be
 \sqrt{-\frac{1}{2}{\cal K}^{ab} {\cal K}_{ab}}\; =\;(1-2\gamma) \left(\frac{2\pi \alpha'}{8\pi G r^2}\right)\;. \label{wtf}
\ee

Imposing  the limit $\;\gamma\rightarrow \varepsilon^2\;$, one can see that our previous results for the undeformed star in \cite{StringFluid} have essentially been restored. More specifically, $\;|\mathcal{K}_{10}|\to\frac{\alpha'}{4 G r^2}\;$, $\;	|\mathcal{K}_{23}|^2\to\mathcal{O}(\varepsilon^2)\;$ and $\;|\mathcal{K}_{T0}|^2\to\mathcal{O}(\varepsilon^2)\;$.

The solution of $|\mathcal{K}_{10}|^2$ corresponds to the existence of an electric, point-like charge at the center of the star, for which
		\begin{equation}
			|q_{core}|\;=\;\frac{\alpha'}{4G} \left( 1-\frac{5}{2}\gamma \right)\;. \label{qe}
		\end{equation}
Similarly, having a non-vanishing $\sqrt{\mathcal{K}^{23}\mathcal{K}_{23}}$ corresponds to the existence of a magnetic, point-like charge (or monopole) at the center of the star, whereby
		\begin{equation}
			|q^{core}_m|\;=\;\frac{\alpha'}{4G}\sqrt{\gamma}\;.
		\label{qm}
		\end{equation}

As already discussed, the total charge (electric plus magnetic) of the star needs to vanish, as its energy density and, thus, its electric and magnetic fields all have to  vanish in the exterior. So that there has to be a charge $\;q_{out}=-q_{core}\;$ that is spread out uniformly over the star’s outer surface and, likewise, there is a uniformly  spread magnetic charge $\;q^{out}_m=-q^{core}_m\;$ on the
same surface.

		\section{Dirac strings}

		Here, we briefly review the origin of Dirac strings in the presence of magnetic monopoles  \cite{Dirac1} (see \cite{Hera} for a modern take) and adapt  this to the defrosted star framework as the discussion proceeds.

		In the single-monopole picture, one first identifies that there is  no finite vector potential that  can be used to describe a
	magnetic induction of the form $\;\vec{B}=\frac{q_m}{r^2}\hat{r}\;$
		throughout the spatial manifold. The best that one can do is to take a combination of vector potentials such  that each one, individually, blows up along  one half of the polar axis. For instance,
		\begin{eqnarray}
			\vec{A}^{\bullet}&=&q_m\frac{1-\cos\theta}{r\sin\theta}\hat{\phi}\;,
			\nonumber \\
			\vec{A}&=&-q_m\frac{1+\cos\theta}{r\sin\theta}\hat{\phi}\;,
		\end{eqnarray}
		as each reproduce the desired $\vec{B}$ field but diverge
		on the negative ($\theta=\pi$) and positive  ($\theta=0$)
		half of the $z$ axis, respectively.
		Importantly, the two potentials are related by a gauge transformation that respects the Coulomb gauge (as appropriate for a stationary configuration):
		$\;\vec{A}^{\bullet}=\vec{A} + \vec{\nabla}\Lambda\;$, where
		$\;\Lambda=2 q_m \phi\;$.

		The way around this is to add to either vector potential a magnetic induction --- that of a Dirac string ---   which conspires
		to cancel off the divergent contribution along the half axis and, at the same time, restores the relevant  Bianchi identity, $\;\vec{\nabla}\cdot\vec{B}=0\;$. These are, respectively,
		 \begin{equation}
			\begin{split}
				&\vec{B}_{Dirac}^{\bullet}(z<0)\;=\;4\pi q_m\delta(x)\delta(y)\Theta(-z)\hat{z}\;,\\
				&\vec{B}_{Dirac}(z>0)\;=\ -4\pi q_m\delta(x)\delta(y)\Theta(z)\hat{z}\;,
			\end{split}
		\end{equation}
         with the  caveat that neither of these fields can be written directly in terms of the curl of a vector potential.
         Notice that either string extends out to infinity and, furthermore,
         that the choice of which string (and its accompanying potential) is irrelevant
         inasmuch as it depends solely on one's choice of gauge.

         Now, for our case, we rather have two sources of magnetic charge to consider. One is the monopole at the center of the star, $q^{core}_m$, and the other is the magnetic charge $\;q^{out}_m =-q^{core}_m\;$ that is uniformly distributed over the outer surface at $\;r=R\;$. For calculational purposes, because of spherical symmetry
         and the magnetic analogue of Gauss' law, the latter can be regarded as a point charge
         sitting at the center of the star but such that its effects
         are only felt for the region $\;r\geq R\;$.
        Accordingly, there is  a net magnetic charge  of zero for the region outside of
        the star and one of $q_m^{core}$ for the interior.
        As such, the magnetic induction $\;\vec{B}=\frac{q_m}{r^2}\hat{r}\;$  must
        terminate at the star's surface and the contributions from
        the Dirac strings must do so as well. The latter is accomplished  with
        the superimposed forms
        \begin{equation}
			\begin{split}
				&\vec{B}^{\bullet}_{Dirac}(z<0)\;=\; 4\pi q^{core}_m\delta(x)\delta(y)\left[\Theta(-z)-  \Theta(-z-R) \right]\hat{z}\;,\\
				&\vec{B}_{Dirac}(z>0)\;=\; - 4\pi q_m^{core}\delta(x)\delta(y)\left[ \Theta(z)-  \Theta(z-R) \right]\hat{z}\;.
			\end{split}\label{mag}
		\end{equation}
It is easy to check that these expressions vanish at $\;r=R\;$, as
they must for consistency.

Thus, the Dirac string is ``trapped'' between the two sources, not unlike the electric  flux  lines in the frozen star setup.  The situation is depicted schematically  in Figure~1
for the $\;z=0\;$ case.
		\begin{figure}[H]
			\centering
			\includegraphics[width=0.8\textwidth]{
			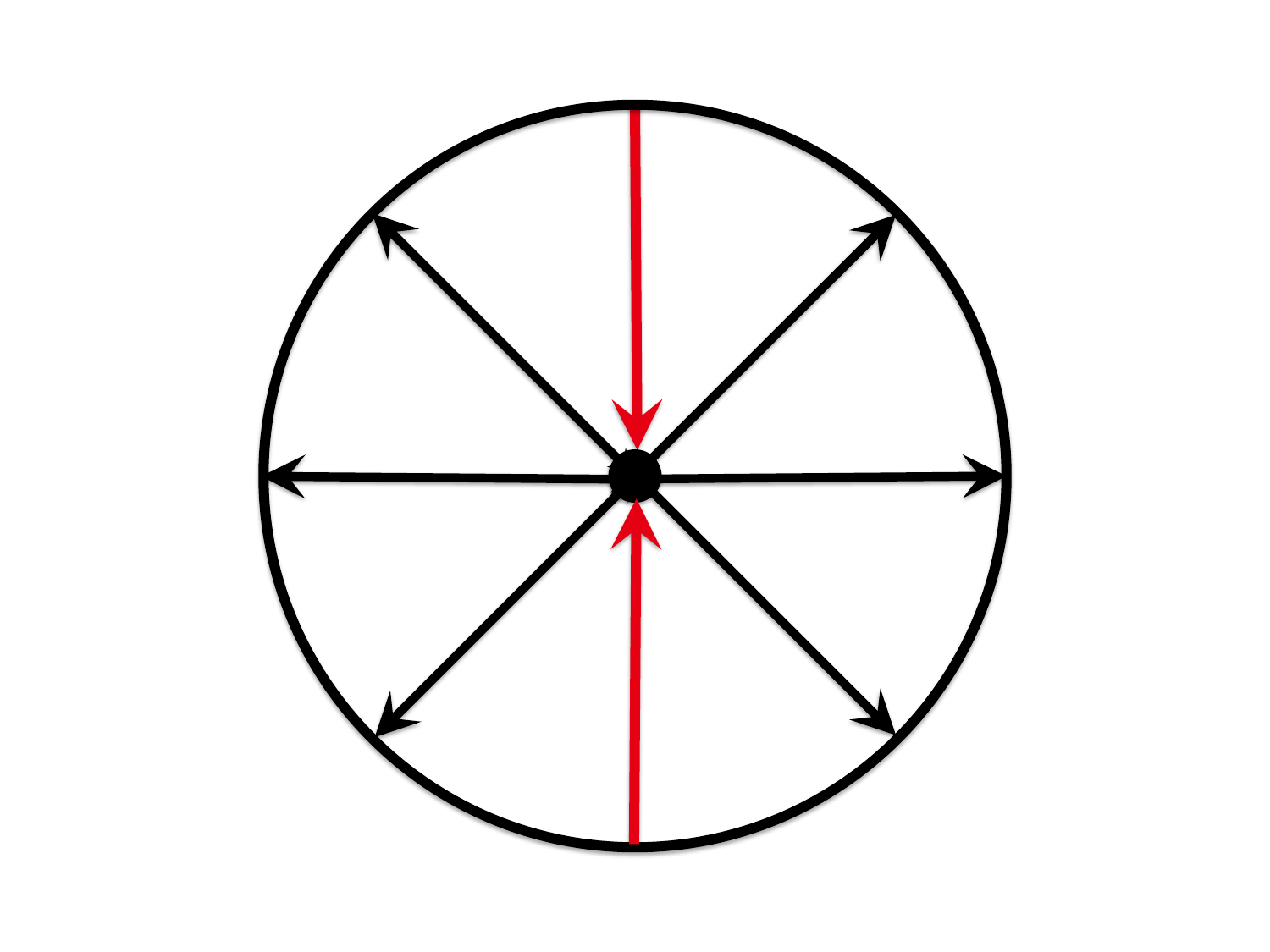}
\caption{A magnetic monopole enclosed by a shell of equal and opposite magnetic charge. The two red arrows are meant to be superimposed along the positive $z$-axis, representing two finite-length Dirac strings as in Eq.~(\ref{mag}).}
			\label{fig1}
		\end{figure}

The appearance of both magnetic and electric charges in the defrosted star solution might have suggested that there should be some constraint arising from a Dirac-like quantization condition involving both charges. However, from an exterior perspective, the star is completely neutral and  there are no free magnetic monopoles or electrically charged particles available to impose such a condition
on. Hence,  Dirac quantization is not relevant to our discussion.

A final point to consider here is about the stability of this dyonic configuration. As previously mentioned, the fixing of the mass by the boundary conditions of the gravitational field at infinity prevents that star from changing its radius. This can modeled by an effective repulsive  force due to a Lagrange-multiplier term in the Lagrangian, which  is then  responsible for cancelling off the attractive electrostatic force between the charges at the core and  outer surface \cite{StringFluid}. The essence of the calculation is that one varies the Lagrange-multiplier term by $A_{a}$, which then, by virtue of the Gauss'-law constraint  at the outer surface $\;\vec{\nabla}\cdot\vec{D}=q^{out}_e  \delta (\vec{r}-R\hat{r})\;$, leads to an expression for the effective force that is equal and opposite to the electrostatic force. This argument seems problematic for the magnetic sector given that the magnetic analogue to the Gauss' law constraint,  $\;\vec{\nabla}\cdot\vec{B}=0\;$, is identically  vanishing. However, the mass and charges are fixed by the boundary conditions leading to the fixing of the radius of the defrosted star. Hence, the magnetic force between the opposite charges is also canceled.


		\section{Perturbations}

		Here,  our focus is on obtaining the linearized equations of motion, $\;\delta G^{a}_{\;\;b}=8\pi G ~\delta T^a_{\;\; b}\;$, by perturbing  the  Lagrangian, which includes the Lagrangians of  Einstein--Hilbert and  Born--Infeld along with tachyon matter as prescribed in the defrosted star framework. The ultimate objective in this section  is to verify the wave equation and thus its solutions from \cite{Tom1}. In this context, we also treat  $\gamma$ as a perturbative parameter and expand the linearized equations to leading order in $\gamma$. The $\varepsilon^2$ corrections are treated as negligible.

		\subsection{ Einstein tensor perturbations}

To start, metric perturbations are expanded in spherical harmonics and simplified via  the Regge--Wheeler gauge \cite{mantle}. We restrict to    even-parity perturbations, as this is what  directly  couples to the density and pressure fluctuations.
The first-order perturbed metric goes as  \cite{lionspaw,ross}
\begin{equation}
	\resizebox{1.2\textwidth}{!}{$
		\begin{pmatrix}
			-\gamma\dfrac{r^2 }{R^2}  \left[ 1 +  Y_{\ell m} e^{i\omega t} H_0(r) \right] & -i\omega H_1(r)  Y_{\ell m} e^{i\omega t} & 0 & 0 \\
			-i\omega H_1(r)  Y_{\ell m} e^{i\omega t} & \gamma^{-1} \left[ 1 -  Y_{\ell m} e^{i\omega t} H_0(r) \right] & 0 & 0 \\
			0 & 0 & r^2 \left[ 1 -  Y_{\ell m} e^{i\omega t} K(r) \right] & 0 \\
			0 & 0 & 0 & r^2 \sin^2\theta \left[ 1 -  Y_{\ell m} e^{i\omega t} K(r) \right]
		\end{pmatrix} \;\cdot \hspace{1in}
		$}
\end{equation}
Here, the spherical harmonics really indicate a sum of these over all $\ell$ and $m$ values. The  three radial functions characterizing the even-parity metric perturbations are	$H_0(r), H_1(r), K(r)$. The radial dependence will often  be omitted for brevity.

The Einstein equations for the perturbations constitute a set of dynamical evolution equations containing second-order time derivatives of the fluctuations and, in addition, a set of initial-value  (or constraint) equations involving only first-order time derivatives.

The perturbations of the key Einstein-tensor elements are found to be as follows \cite{Tom1} (with a single spherical harmonic again implying a sum thereof):
\bea
&& -\delta\rho\;=\;\delta G^t_{\;\;t}\;=\;e^{i \omega t} Y_{\ell m} ~\frac{1}{r^2} \times \cr &&
\cr &&
\left[ \frac{1}{2}(\ell^2+\ell-2)K +\frac{1}{2}~\ell(\ell+1) H_0 -\gamma (r^2 K''+3rK')+\gamma( r H'_0+H_0)\right]\;,
\label{drho} \\
&& \cr
&& \delta p \;=\;\delta G^r_{\;\;r} \;=\; e^{i \omega t} Y_{\ell m} ~\frac{1}{r^2}\times\cr &&
\cr &&
\left[ \frac{1}{2}(\ell^2+\ell-2)K -\frac{1}{2}~\ell(\ell+1) H_0-\gamma \widetilde\omega^2 K -2\gamma  \left( r K'-\gamma \widetilde\omega ^2\frac{H_1}{ r}\right)
+\gamma(r H'_0+3 H_0)
\right] \;. \hspace{0.5in}
\label{dpi}
\eea
Anticipating the scaling $\omega\sim \gamma$, we have redefined, as in \cite{Tom1},
$\;R^2\omega^2=\gamma^2 \widetilde{\omega}^2\;$.

Let us  next consider one of the initial-value equations,
$\delta G^r_{\;\;\theta}\; = 8\pi G \delta T^r_{\;\;\theta}\;$.  As for the Einstein tensor perturbation
\begin{equation}
\delta G^r_{\;\;\theta}\; =\;e^{i \omega t}\partial_\theta Y_{\ell m} ~ \frac{1}{4 r} \times \left[2 \gamma\left(  r K'-\gamma\widetilde{\omega}^2\frac{H_1} { r}\right)
-2\gamma\left(r H_0'+2 H_0\right)\right]\;,
		\end{equation}
this can be set to vanish by a choice of initial conditions, which we choose to impose,
\be
2 \gamma\left(  r K'-\gamma\widetilde{\omega}^2\frac{H_1} { r}\right)
-2\gamma\left(r H_0'+2 H_0\right)\;=\;0\;.
\label{const-theta-r}
\ee
Note that, in spite of the appearance of  $\widetilde{\omega}^2$,  this equation involves only a first-order time derivative because the metric perturbation $\;\delta g_{tr} \sim \omega H_1\;$ is already dressed by a factor of $\omega$.

To proceed, we recall that, to leading order in $\gamma$, $\;\delta p + \delta \rho =0\;$. Then, because  of $\;\delta p,\ \delta \rho \sim \gamma\;$, it follows that  $\;\delta p + \delta \rho \sim \gamma^2\;$.
Now consider that,  from Eqs.~(\ref{drho}) and~(\ref{dpi}),
\bea
&& \delta G^r_{\;\;r} - \delta G^t_{\;\;t} \;=\;e^{i \omega t} Y_{\ell m} ~\frac{1}{r^2}\times
 \cr &&
\cr &&
\left[-\ell(\ell+1)H_0-2 \gamma\left(  r K'-\gamma\widetilde{\omega}^2\frac{H_1} { r}\right)
+ \gamma  \left(r^2 K''+3rK'-\widetilde\omega^2 K\right)+2\gamma H_0\right]\;. \hspace{0.5in}
\label{dp-Plus-drho}
\eea

The conclusion is then that $\;H_0 \sim \gamma^2\;$ and, to leading order in $\gamma$,
\be
\delta p \;= \;-\delta \rho \;=\; e^{i \omega t} Y_{\ell m} (\ell^2+\ell-2)\frac{K}{2r^2}\;.
\label{solGt}
\ee
It follows that, as expected, $\;K\sim\gamma\;$.
Furthermore, Eq.~(\ref{const-theta-r}) now implies  that, to leading order,
\be
r K'-\gamma\widetilde{\omega}^2\frac{H_1} { r}\;=\;0\;.
\label{solGtr}
\ee

We may now identify the leading-order (up to $\gamma^2$) expressions for $\delta G^t_{\;\;t}$ and $\delta G^r_{\;\;r}$,
\bea
&& \delta G^t_{\;\;t}\;=\;e^{i \omega t} Y_{\ell m} ~\frac{1}{r^2} \times
\left[ \frac{1}{2}(\ell^2+\ell-2)K +\frac{1}{2}~\ell(\ell+1) H_0
-\gamma (r^2 K''+3rK')\right]\;, \hspace{0.5in}
\label{drhoLeading} \\
&& \cr
&& \delta G^r_{\;\;r} \;=\; e^{i \omega t} Y_{\ell m} ~\frac{1}{r^2}\times
\left[ \frac{1}{2}(\ell^2+\ell-2)K -\frac{1}{2}~\ell(\ell+1) H_0-\gamma \widetilde\omega^2 K
\right] \;.
\label{dpiLeading}
\eea

For future reference, we also calculate $\;\delta G^r_{\;\;r} + \delta G^t_{\;\;t}=\delta p -\delta\rho\;$,
which, to order $\gamma^2$ and through the use of Eqs.~(\ref{drhoLeading}) and (\ref{dpiLeading}), simplifies to
\bea
&& \delta G^r_{\;\;r} + \delta G^t_{\;\;t} \;=\;e^{i \omega t} Y_{\ell m} ~\frac{1}{r^2}\times
\left[(\ell^2+\ell-2)K -\gamma  \left(r^2 K''+3rK'+\widetilde\omega^2 K\right)\right]\;,\hspace{0.5in}
\label{dp-Minus-drho}
\eea
with corrections to this equation being  of order $\gamma^3$.

As for the remaining perturbed elements of the Einstein tensor,
these do not contribute  additional information to the leading-order equation of motion for $K$.
One can use them, however,  to find the    higher-order corrections.

\subsection{Energy-momentum tensor perturbations}

In  this  subsection, the factors of $e^{i\omega t}Y_{\ell m}$
that accompany the relevant expressions for $\delta T^a_{\;\;b}$ and some other perturbed quantities
will be dropped, and we also set $\;2\pi\alpha'=1\;$.

Let us start with the calculation of
\be
\delta T^a_{\ b}\;=\;\delta \left(\frac{{\cal K}^{a c}{\cal K}_{b c}}{\sqrt{-\frac{1}{2} {\cal{K}}^2}}\right)\;.
\ee
As in the calculation of $\delta G^a_{\ b}$, only the
leading-order terms will be retained.

First consider that
\be
\delta T^t_{\ t}\;= \;\delta \frac{K^{tr}K_{tr}}{\sqrt{-K^{tr}K_{tr}}}\;.
\ee
Recalling that
$\;K^{tr}K_{tr}=- D^i D_i\;$
and
$\;\sqrt{-K^{tr}K_{tr}}=\sqrt{D^i D_i}\;$, we denote
$-D^iD_i$ as $-D^2$
and
$\sqrt{D^i D_i}$ as $D$.~\footnote{We are assuming $\;D_i \geq 0\;$
without loss of generality.} So that,  to leading order,
\be
{\sqrt{-\frac{1}{2} {\cal{K}}^2}}\;=\;D\;,
\label{sqrtK}
\ee
\be
-\delta T^t_{\ t}\;=\; \delta \rho \;=\;  \delta D\;,
\label{solTt}
\ee
and similarly, to leading order,
\be
\delta T^r_{\ r} \;=\; \delta p \;=\; -\delta D\;.
\label{ppppp}
\ee

Additionally, let us calculate
\bea
\delta p +\delta\rho\;=\;\delta T^r_{\ r} -\delta T^t_{\ t}\;=\;
\delta\left(- \frac{D^2}{\sqrt{-\frac{1}{2} {\cal{K}}^2}}\right)+
\delta\left( \frac{D^2 +|{\cal{K}}_{T0}|^2}{\sqrt{-\frac{1}{2} {\cal{K}}^2}}\right)
\;=\;
\delta\left( \frac{|{\cal{K}}_{T0}|^2}{\sqrt{-\frac{1}{2} {\cal{K}}^2}}\right)\;. \nonumber \\
\label{Minus-rp1}
\eea

As we will show next, to leading order
\be
 \delta\left( \frac{|{\cal{K}}_{T0}|^2}{D}\right)\;=\; -\frac{|{\cal{K}}_{T0}|^2}{D^2}
\delta D(r) \;=\;
- 2\gamma \delta D(r)\;.
\label{solsumrp}
\ee

To justify Eq.~(\ref{solsumrp}), let us begin with
\bea
\delta\left(\frac{|{\cal{K}}_{T0}|^2}{\sqrt{-\frac{1}{2} {\cal{K}}^2}}\right)
\;=\;\frac{K^{T0}\delta K_{T0} }{\sqrt{-\frac{1}{2} {\cal{K}}^2}}+|{\cal{K}}_{T0}|^2
\delta \frac{1}{\sqrt{-\frac{1}{2} {\cal{K}}^2}}\;.
\eea
As our interest is in the leading-order result, we can ignore the $r$--$t$ sector metric perturbation $\;H_0\sim \gamma^2\;$ and, because $\;|{\cal{K}}_{T0}|^2 \sim \gamma\;$,   approximate $\;\sqrt{-\frac{1}{2} {\cal{K}}^2}=D\;$ in the second term on
the right. This
yields
\bea
\delta\left( \frac{|{\cal{K}}_{T0}|^2}{\sqrt{-\frac{1}{2} {\cal{K}}^2}}\right)
\;=\;\frac{2K^{0T}\delta K_{0T} }{D}+|{\cal{K}}_{T0}|^2
\delta \frac{1}{D}\;.
\eea

Then, because $\;K_{0T}=\partial_t T\;$ and the background quantities are time-independent, we may integrate by parts,
\be
K^{0T}\delta K_{0T}\;=\;\partial^t T ~\delta(\partial_t T)\;=\;-\delta T~ \Box T\;,
\ee
where the right-most  result follows from the homogeneity of the
tachyon. Meanwhile, the tachyon equation of motion is $\;\Box T=0\;$ because the tachyon is at the minimum of its potential in this
background. This leaves the
second term, which  can also be evaluated,
\bea
|{\cal{K}}_{T0}|^2 \delta \frac{1}{D}\;=\;-|{\cal{K}}_{T0}|^2 \delta{D} \frac{1}{D^2}\;=\;
-2\gamma \delta D(r)\;.
\eea

To summarize, we found that to leading order, which is in this case $O(\gamma^2)$,
\be
\delta p +\delta\rho\;=-\;2\gamma \delta D\;=\;- 2 \gamma\delta \rho\;.
\label{TttTrr}
\ee

To conclude this subsection, we calculate $\delta p -\delta\rho$ using Eqs.~(\ref{ppppp}) and~(\ref{TttTrr}),
\bea
\delta p -\delta\rho &=& 2\delta p -(\delta p+\delta\rho)\nonumber \\
&=& -2\delta D + 2\gamma \delta D\;.
\label{dp-minus-drho}
\eea

Importantly to what follows, $\delta p$ receives no further
corrections due to fluctuations of the tachyon and magnetic fields.
The former because of the application of the tachyon equation of motion as discussed above and the latter because the transverse pressure  $\;q= \frac{|K_{23}|^2}{D}\;$ should be regarded as a fixed quantity,
as is standard practice, in the calculation of $\delta p$.~\footnote{The same constraint of fixed  $q$ does not apply to $\delta\rho$
because $\rho$ only appears in tandem with $p$ in the conservation equation and $\;p+\rho= 2 q\;$.}
Hence,
$\;\delta p= -\delta D\;$ to at least order $\gamma^2$.

\subsection{Perturbed Einstein equations}

Our goal here is to derive the dynamical equation for the mode $K$ from the Einstein equation
$\;\delta G^r_{\;\;r}+\delta G^t_{\;\;t}=\delta p -\delta\rho\;$,
where we now set $\;8\pi G = 1\;$.

Let us recall the leading-order result  Eq.~(\ref{solGt}) and compare it to the leading-order result in  Eq.~(\ref{dpiLeading}).
Then, since $\;\delta p=-\delta D\;$, the conclusion is that, to order $\gamma$,
\be
\;\delta D\;=\;-e^{i \omega t} Y_{\ell m}\frac{1}{r^2}\left(\frac{1}{2}(\ell^2+\ell-2)K-\frac{1}{2}~\ell(\ell+1) H_0-\gamma \widetilde\omega^2 K\right)\;.
\ee

We cannot be conclusive at this point as to what constitutes
the $\gamma^2$-order  correction to $D$.
However, if we impose the condition that
\be
\frac{1}{2}~\ell(\ell+1) H_0\;=\;-\gamma \widetilde\omega^2 K\;
\label{H0cond}
\ee
and recall that $\;\delta p=-\delta D\;$ is valid to at least order
$\gamma^2$, then it follows that
\be
\;\delta D\;=\;-e^{i \omega t} Y_{\ell m}\frac{1}{r^2}\left(\frac{1}{2}(\ell^2+\ell-2)K\right)\; +\; {\cal O}[\gamma^3]\;.
\label{dDF}
\ee
This condition guarantees, as shown in the Appendix, that to order $\gamma^2$, the perturbations obey the same equation of state as that of the lowest-order term  in Eq.~(\ref{TttTrr}),
\be
\delta{\rho}+\delta p \;=\;-2\gamma \delta\rho\;,
\label{dpdrhogamma2}
\ee
in a way that is consistent with the final equation for $K$, Eq~(\ref{efadoodle})
.

It now follows from Eq.~(\ref{dp-minus-drho}) that
\be
\delta p -\delta\rho =
 e^{i \omega t} Y_{\ell m}\frac{1}{r^2}\left[(\ell^2+\ell-2) K
- \gamma (\ell^2+\ell-2)K\right]\;.
\label{p-rhoF}
\ee

We are finally  ready to write the Einstein equation
$\delta G^r_{\;\;r}+\delta G^t_{\;\;t}=\delta p -\delta\rho$ by combining Eqs.~(\ref{dp-Minus-drho}) and (\ref{p-rhoF}),
\bea
 (\ell^2+\ell-2)  K- \gamma \left(r^2 K''+3rK'+\widetilde\omega^2K\right) \;=\;  (\ell^2+\ell-2)K-\gamma(\ell^2+\ell-2)K\;, \ \ \ \
\label{F2}
\eea
which simplifies to
\be
r^2 K''+3 r K' +\left[\widetilde{\omega}^2-(\ell^2+\ell-2)\right]K\;=\; 0\;.
\label{efadoodle}
\ee

This final equation  is  the same as the wave equation which was obtained in \cite{Tom1}, Eq.~(4.7).  As such, all of the previous results from \cite{Tom1} regarding the solution of Eq.~(\ref{efadoodle})
persist. Most importantly,  the internal sound velocity of
these waves is non-relativistic, scaling linearly with $\gamma$,
and they  slowly leak from the star's interior, at a decay rate that is proportional to $\gamma^2$.

Once the solution for $K$ is known, one can use Eqs.~(\ref{H0cond})
and~(\ref{solGtr}) to solve for $H_0$ and $H_1$, respectively.


		\section{Conclusion}

In this work, we have strengthened and improved our previous results about the spectrum of the non-radial oscillatory modes of the defrosted star model.  Previously, we have shown that the perturbations of a defrosted star give rise to non-relativistic modes with sound velocities of order  $\gamma$ and  parametrically long lifetimes of order $1/\gamma^2$ \cite{fluctuations,Tom1}. The quantum deformation or defrosting parameter, $\gamma$, can be viewed as the analogue of the string coupling in the BH polymer model, whose modes behaved similarly but with $g_s$ in place of $\gamma$ \cite{collision}. Here, we have validated the previous results, but the improvement in the current analysis is that the framework now includes an explicit Lagrangian for the matter source: A Born--Infeld form that is related to Sen's model for $D$-brane decay \cite{SenReview}, with refinements leading to  the  picture of a fluid of open strings carrying electric flux \cite{GHY}. This inclusion puts our results on much firmer conceptual ground and leads to a much simplified set of equations for the matter perturbations. Moreover, armed with this Lagrangian, one can investigate perturbations to configurations that deviate strongly from both spherical symmetry and staticity.  Meaning that the model is well geared for studying out-of-equilibrium scenarios, such as BH merger events. Because of the explicit nature of the Lagrangian, any results from studies along these lines can be readily compared to the predictions of standard general relativity.

		A central result of this study is the characterization of the defrosted  star as a dyonic object.  Whereas the ultrastable frozen star carries only electric charges,
		the inclusion of a magnetic monopole at the star's center, as well as an equal but oppositely charged layer at the outer surface, was shown to be a necessary ingredient
		for the purpose of  deviating away from
		the ultrastability condition of $\;p+\rho=0\;$ while preserving spherical symmetry. An interesting side feature
		of this setup is the presence of a truncated Dirac string,
		which is consistent with both the external Schwarzschild vacuum
		and Dirac's self-consistent formalization of a monopole.

	    As emphasized in previous papers,  the resulting oscillation spectrum offers an observationally relevant  feature  of frozen stars with a distinctive signature that is in direct contrast to  the rapidly decaying quasi-normal modes of classical BHs, and possibly differing from  the modes  of other classes of BH mimickers as well. This feature may then  serve as a potential discriminator in the analysis of the empirical data from future gravitational-wave
	    experiments.

\section*{Acknowledgments}
We thank Tom Shindelman for sharing her code for calculating the perturbations and Frans Pretorius for emphasizing the need for a Lagrangian formulation of the equations of motion and the perturbation equations.
The research is supported by the German Research Foundation through a German-Israeli Project Cooperation (DIP) grant ``Holography and the Swampland'' and by VATAT (Israel planning and budgeting committee) grant for supporting theoretical high energy physics.
The research of AJMM received support from an NRF Evaluation and Rating Grant 119411. AJMM thanks Ben Gurion University for their hospitality on past visits. RB thanks the theory department at CERN, the department of theoretical physics at the University of Geneva and the Arnold Sommerfeld Centre at LMU, Munich, for their hospitality.

\begin{appendices}

    \section{Justifying Eq.~(\ref{H0cond})}

Equation~(\ref{H0cond}) implies, as does Eq.~(\ref{dDF}) via $\;\delta p=-\delta D\;$, that  the order-$\gamma^2$ term  in $\delta p$, which we now denote by $\delta p^{(2)}$, vanishes,
\be
\delta p^{(2)}\;=\;0\;.
\ee

Meanwhile, from Eq.~(\ref{drhoLeading}), the order-$\gamma^2$
term in $\delta\rho$, which is similarly  denoted by $\delta\rho^{(2)}$, is given by
\be
-\delta\rho^{(2)}\;=\;e^{i \omega t} Y_{\ell m} ~\frac{1}{r^2}\left[\frac{1}{2}~\ell(\ell+1) H_0 -\gamma (r^2 K''+3rK')\right]\;,
\ee
which, via the application of  Eq.~(\ref{H0cond}), can be rewritten as
\be
\delta\rho^{(2)}\;=\;e^{i \omega t} Y_{\ell m} ~\frac{1}{r^2}
\left[ \gamma (r^2 K''+3rK'+\widetilde\omega^2 K)\right]\;.
\ee

It follows that
\be
 \delta\rho^{(2)}+\delta p^{(2)} \;=\; \delta \rho^{(2)}\;=\;
\;e^{i \omega t} Y_{\ell m} ~\frac{1}{r^2}
\left[\gamma (r^2 K''+3rK'+\widetilde\omega^2 K)\right]\;.
\ee

Requiring  that the sum $\delta\rho^{(2)}+\delta p^{(2)}$
matches the leading-order (or $\gamma^2$ order) term on the right-hand side of Eq.~(\ref{dpdrhogamma2}), we can then use
$\;\delta \rho^{(1)} = \delta D^{(1)}\;$ and Eq.~(\ref{dDF}) for $\delta D^{(1)}$  to obtain
\be
\gamma (r^2 K''+3rK'+\widetilde\omega^2 K)\;=\;\gamma~(\ell^2+\ell-2) K
\;,
\ee
which is the same as the final  equation for $K$, Eq.~(\ref{efadoodle}).

\end{appendices}

			
		\end{document}